\documentclass[floats,floatfix,amssymb,prd,twocolumn,superscriptaddress,nofootinbib, preprintnumbers]{revtex4-1}
	
\usepackage{subcaption}
\usepackage{ragged2e}
\DeclareCaptionJustification{justified}{\justifying}
\usepackage{amssymb,amsmath,verbatim,mathtools,needspace,enumitem,etoolbox,graphicx,physics,microtype,afterpage,bm}
\usepackage{esint}
\usepackage[dvipsnames, usenames]{xcolor}
\definecolor{linkcolor}{rgb}{0.0,0.3,0.5}
\usepackage[unicode, colorlinks=true, linkcolor=linkcolor, citecolor=linkcolor, filecolor=linkcolor,urlcolor=linkcolor, pdfusetitle]{hyperref}
\usepackage[all]{hypcap}
\usepackage[T1]{fontenc}
\usepackage[utf8]{inputenc}
\usepackage{tabularx}
\usepackage{float}
\newcommand\underrel[3][]{\mathrel{\mathop{#3}\limits_{%
      \ifx c#1\relax\mathclap{#2}\else#2\fi}}}
\usepackage{soul}
\usepackage{lmodern}
\allowdisplaybreaks
\usepackage{tikz}
\usepackage{color}
\usepackage{framed}
\usepackage{hyperref}
\hypersetup{colorlinks, citecolor=bluscuro, linkcolor=black, urlcolor=bluscuro}
\definecolor{rossos}{cmyk}{0,1,1,0.55}
\definecolor{bluscuro}{rgb}{0.15, 0.2, .85}
\definecolor{bluchiaro}{cmyk}{1,.3,0.,0.1}
\definecolor{ForestGreen}{rgb}{0.13, 0.55, 0.13}

\newcommand{\be}{\begin{equation}}
\newcommand{\ee}{\end{equation}}

\def\BH{\text{\tiny BH}}
\newcommand{\llp}{\left [}
\newcommand{\rrp}{\right ]}
\newcommand{\lp}{\left (}
\newcommand{\rp}{\right )}

\def\lsim{\mathrel{\rlap{\lower4pt\hbox{\hskip0.5pt$\sim$}}
    \raise1pt\hbox{$<$}}}         
\def\gsim{\mathrel{\rlap{\lower4pt\hbox{\hskip0.5pt$\sim$}}
    \raise1pt\hbox{$>$}}}         

\makeatletter
\newcommand{\subsetsim}{\mathrel{\mathpalette\subset@sim\relax}}
\newcommand{\subset@sim}[2]{%
  \vtop{\offinterlineskip\m@th
    \ialign{\hfil##\cr
     ~$#1\subset$\cr\noalign{\kern0.5pt}\scalebox{0.9}{$#1\sim$}\cr
    }%
  }%
}
\makeatother

\makeatletter
\def\l@subsubsection#1#2{}
\makeatother

\newcommand{\sapienza}{Dipartimento di Fisica, Sapienza Università 
	di Roma, Piazzale Aldo Moro 5, 00185, Roma, Italy}
\newcommand{\infn}{INFN, Sezione di Roma, Piazzale Aldo Moro 2, 00185, Roma, Italy}

\begin{document}

\title{Tidal deformability of black holes surrounded by thin accretion disks}

\author{Enrico Cannizzaro}
\email{enrico.cannizzaro@tecnico.ulisboa.pt}
\affiliation{CENTRA, Departamento de Física, Instituto Superior Técnico – IST,
Universidade de Lisboa – UL, Avenida Rovisco Pais 1, 1049 Lisboa, Portugal}

\author{Valerio De Luca}
\email{vdeluca@sas.upenn.edu}
\affiliation{Center for Particle Cosmology, Department of Physics and Astronomy,
University of Pennsylvania 209 South 33rd Street, Philadelphia, Pennsylvania 19104, USA}

\author{Paolo Pani}
\email{paolo.pani@uniroma1.it}
\affiliation{\sapienza}
\affiliation{\infn}


\begin{abstract}
\noindent
The tidal Love numbers of self-gravitating compact objects describe their response to external tidal perturbations, such as those from a companion in a binary system, offering valuable insights into their internal structure. For static tidal fields, 
asymptotically flat black holes in vacuum exhibit vanishing Love numbers in general relativity, even though this property is sensitive to the presence of an external environment. In this work we study the tidal deformability of black holes surrounded by thin accretion disks, showing that the Love numbers could be large enough to mask any effect of modified gravity and to intrinsically limit tidal tests of black-hole mimickers. Furthermore, we investigate the measurability of the tidal parameters with next-generation gravitational wave experiments, like LISA and Einstein Telescope. Our findings suggest that these parameters could be measured with high precision, providing a powerful tool to probe the environment around coalescing binary systems.
\end{abstract}

\preprint{ET-0481A-24}
\maketitle

\section{Introduction}
\label{sec:intro}
\noindent
The detection of gravitational waves (GWs) from merging binary systems heralds a new era in the study of gravity and compact objects in the strong-field regime~\cite{Bailes:2021tot}. As the next generation of detectors, such as LISA and the Einstein Telescope (ET), comes online, a significant increase in observations and the discovery of novel GW sources are expected~\cite{Punturo:2010zz, Sathyaprakash:2019yqt, Maggiore:2019uih, Reitze:2019iox, Kalogera:2021bya,Branchesi:2023mws,LISA:2017pwj,Colpi:2024xhw}. This necessitates the development of highly accurate waveform templates based on a thorough understanding of the conservative and dissipative dynamics of two-body systems~\cite{Chia:2023tle, Chia:2024bwc,LISAConsortiumWaveformWorkingGroup:2023arg}. Tidal effects are a critical component in the inspiral of binary systems. By measuring these effects through GW observations, we can gain valuable insights into the interior structure and environments of compact objects. For example, tidal measurements can provide information about the equations of state of neutron stars (see Refs.~\cite{GuerraChaves:2019foa, Chatziioannou:2020pqz} for reviews), reveal the existence of exotic compact objects~\cite{Cardoso:2017cfl, Cardoso:2019rvt, Herdeiro:2020kba, Chen:2023vet, Berti:2024moe}, and potentially uncover new physics at the event horizon scale of black holes (BHs)~\cite{Maselli:2018fay, Datta:2021hvm}.

A persistent analytical framework for characterizing tidal effects involves a set of coefficients called tidal Love numbers (TLNs), which describe the linear conservative response of self-gravitating bodies~\cite{1909MNRAS..69..476L}. Originally formulated within Newtonian gravity, TLNs have been further generalised to a fully relativistic context~\cite{Hinderer:2007mb, Binnington:2009bb, Damour:2009vw}. Within this framework, it is widely recognized that the static TLNs of various families of asymptotically flat BHs in four-dimensional space-times are precisely zero~\cite{Deruelle:1984hq,Binnington:2009bb, Damour:2009vw, Damour:2009va, Pani:2015hfa, Pani:2015nua, Gurlebeck:2015xpa, Porto:2016zng, LeTiec:2020spy, Chia:2020yla, LeTiec:2020bos, Hui:2020xxx, Charalambous:2021mea, Charalambous:2021kcz, Creci:2021rkz, Bonelli:2021uvf, Ivanov:2022hlo, Charalambous:2022rre, Katagiri:2022vyz, Ivanov:2022qqt, Berens:2022ebl, Bhatt:2023zsy, Sharma:2024hlz, Rai:2024lho}. However, this characteristic is fragile and can be violated in scenarios involving BH mimickers~\cite{Pani:2015tga, Cardoso:2017cfl} and exotic compact objects~\cite{Cardoso:2017cfl,Herdeiro:2020kba,Chen:2023vet,Berti:2024moe}, the presence of a cosmological constant~\cite{Nair:2024mya}, theories of modified gravity~\cite{Cardoso:2017cfl, Cardoso:2018ptl, DeLuca:2022tkm, Barura:2024uog} and higher dimensions~\cite{Kol:2011vg, Cardoso:2019vof, Hui:2020xxx, Rodriguez:2023xjd, Charalambous:2023jgq, Charalambous:2024tdj, Charalambous:2024gpf, Ma:2024few}. Furthermore, recent studies have also discussed the nature of TLNs in the presence of nonlinearities~\cite{DeLuca:2023mio, Riva:2023rcm, Ivanov:2024sds}, or for time-dependent tidal perturbations~\cite{Nair:2022xfm, Saketh:2023bul, Perry:2023wmm, Chakraborty:2023zed, DeLuca:2024ufn}.

The absence of conservative tidal deformations for BHs is usually stated in the context of binaries evolving in the vacuum, i.e., in the absence of an external environment. Its presence drastically changes the nature of tidal interactions, possibly resulting into a nonvanishing TLNs for non-vacuum BHs. Few examples were already investigated in the literature, such as BHs surrounded by clouds of ultralight bosonic fields sourced due to accretion or superradiance~\cite{Baumann:2018vus, DeLuca:2021ite, DeLuca:2022xlz, Brito:2023pyl}, accreting BHs~\cite{Capuano:2024qhv}, or by fluids of matter~\cite{Cardoso:2019upw, Cardoso:2021wlq} and thin matter shells~\cite{Katagiri:2023yzm, DeLuca:2024uju}. Among this category one may consider accretion disks.

Accretion disks around BHs are ubiquitous phenomena in astrophysics, which form when matter, such as plasma, gas and dust, spirals towards the BH event horizon due to gravity~\cite{Page:1974he, Thorne:1974ve, Pringle:1981ds, Abramowicz:1988sp,2002apa..book.....F, Abramowicz:2011xu}. As the material accelerates and loses gravitational potential energy, it heats up and emits radiation across the electromagnetic spectrum, making accretion disks potentially observable across vast distances~\cite{Paczynski:1979rz, Woosley:1993wj}. Understanding their structure, dynamics, and emission mechanisms is essential for deciphering the physics of BHs and GWs~\cite{ Barausse:2014tra}. Various models have been developed to describe thin accretion disks around BHs, such as the Shakura-Sunyaev model~\cite{Shakura:1972te} and its relativistic extension, the Novikov-Thorne model~\cite{NovikovThorne}.

In this work, following the simplified model developed in Refs.~\cite{Kotlarik:2018nbd,Kotlarik:2022spo,Chen:2023akf} describing a static and axisymmetric thin accretion disk around a Schwarzschild BH, we aim to study the tidal deformability of these objects in binary systems, and investigate their observational measurability at future GW experiments like LISA and ET. We will show that these systems naturally get a nonvanishing TLN induced by the external environment, which may be probed by next-generation GW detectors.
In this paper we use geometrical units $c = G = 1$ and mostly-positive signature. 

\section{effective geometry}
\label{sec: geometry}
\noindent
In this section we are going to review the effective geometry describing a nonrotating BH surrounded by a thin accretion disk following the procedure outlined in Refs.~\cite{Kotlarik:2018nbd,Kotlarik:2022spo,Chen:2023akf}.
This approach is based on the existence of a static solution of the Einstein equations, where an axially symmetric accretion disk with typical thickness much smaller than the BH radius, and thus effectively thin, encircles a Schwarzschild BH (dubbed SBH-disk model).\footnote{During the inspiral phase of the BH binary system, one can note that the characteristic accretion timescale, given by the Salpeter time, is much longer than the orbital period. This allows us to treat accretion as an adiabatic process, assume a static background metric and neglect the time dependence of the metric perturbations.} The disk is simplified to be spatially infinite, with a finite total mass $\mathcal{M}_{\rm d}$, and extending from the BH horizon, where no matter is assumed to be present.

The effective metric describing the system reads in Schwarzschild coordinates~\cite{Kotlarik:2022spo,Chen:2023akf}
\begin{align}
    {\rm d}s^2=&-f(r)\textrm{e}^{2\nu_{\text{\tiny disk}}}{\rm d}t^2+\textrm{e}^{2\lambda_{\text{\tiny ext}}-2\nu_{\text{\tiny disk}}}\frac{{\rm d}r^2}{f(r)}\nonumber\\
&+r^2\textrm{e}^{-2\nu_{\text{\tiny disk}}}\left(\textrm{e}^{2\lambda_{\text{\tiny ext}}}{\rm d}\theta^2+\sin^2\theta {\rm d}\varphi^2\right)\,,\label{superposedmetric}
\end{align}
where $f(r) = 1 - r_\BH/r$ describes the usual Schwarzschild metric function, with $r_\BH = 2 m_\BH$. The terms $\nu_{\text{\tiny disk}}$ and $\lambda_\text{\tiny ext}=\lambda_{\text{\tiny disk}}+\lambda_{\text{\tiny int}}$ describe contributions from the disk itself and from nonlinear interactions with the BH. As we will show later, treating the disk as a small perturbation to the BH allows considering only the interaction part $\lambda_{\text{\tiny int}}$ for the second metric function.\footnote{The absence of any effect of rotation or dragging in the model arises from the fact that the disk has a simplified description in terms of two equally counter-rotating pressureless dust streams following circular geodesics~\cite{Kotlarik:2022spo,Chen:2023akf}.}

The disk potential takes the involved form~\cite{1963ApJ...138..385T, 10.1111/j.1365-2966.2009.14803.x}
\begin{equation}
	\nu_\text{\tiny disk}^{(m, n)} = - W^{(m,n)}\sum_{j=0}^{m + n} \mathcal{Q}_j^{(m,n)} \frac{b^j}{r_b^{j+1}} P_j\left(\frac{|z| + b}{r_b} \right) \,,
    \label{VLpotential}
\end{equation}
where the normalization $W^{(m,n)}$ is chosen in such a way that the total mass of the disk is $\mathcal{M}_{\rm d}$, assuming the disk density falls quickly enough both at the horizon and spatial infinity. In particular,
\begin{equation}
    W^{(m,n)} =  (2m+1) \binom{m + n+ 1/2}{n} \mathcal{M}_{\rm d}\,,
\end{equation}
where $\left(\begin{smallmatrix}
     a  \\
     b 
\end{smallmatrix}\right)$ is the binomial coefficient.
The metric coefficient depends on the scale $b$, which describes approximately the location of maximum density of the accretion profile, and on $r_b = \sqrt{\rho^2 + (|z| + b)^2}$, expressed in terms of the Weyl coordinates
\begin{equation}
    \rho = \sqrt{r (r-2m_\BH)} \sin\theta \,, \quad z = (r-m_\BH) \cos\theta \,.
\end{equation}
Moreover, $P_j$ are the Legendre polynomials, while the coefficients $\mathcal{Q}_j^{(m,n)}$ are defined to be
\[   
\mathcal{Q}_j^{(m,n)} = 
	\begin{cases}
		\sum_{k=0}^n (-1)^k \binom{n}{k} \frac{2^{j-k-m} (2m + 2k -j)!}{(m + k - j)!(2m + 2k + 1)!!} & \text{ if } j\leq m \\
		\sum_{k=j}^{m+n} (-1)^{k-m} \binom{n}{k-m} \frac{2^{j-k} (2k - j)!}{(k-j)!(2k + 1)!!} & \text{ if } j > m \,.
	\end{cases}
\]
Finally, the potential depends on the set of coefficients $(m,n)$, whose values dictate the location and width of the peak. 
For simplicity, in the next section we will fix them to the characteristic values $m = 0, n = 1$, and leave only $b$ as the relevant parameter determining the effective shape of the system's metric. A dedicated discussion on the form of the metric by varying these coefficients can be found in Refs.~\cite{Kotlarik:2022spo,Chen:2023akf}.

Since astrophysical BHs are expected to have a typical mass $m_\BH$ much larger than the mass of the accretion disk $\mathcal{M}_{\rm d}$, it is natural to set $\epsilon=\mathcal{M}_{\rm d}/m_\BH \ll 1$ and consider terms up to $\mathcal{O}(\mathcal{M}_{\rm d}/m_\BH)$. This approximation allows us to simplify $\lambda_\text{\tiny ext}\approx\lambda_{\text{\tiny int}}$, since $\lambda_\text{\tiny disk}$ is quadratic in $\epsilon$~\cite{Kotlarik:2022spo,Chen:2023akf}.
The interaction part $\lambda_\text{\tiny int}$ of the metric function is then found to satisfy the following recurrence relations
\begin{align}
    &\lambda^{(0,0)}_\text{\tiny int} = - \frac{\mathcal{M}_{\rm d}}{r_b} \left( \frac{R_+}{b+m_\BH} - \frac{R_-}{b-m_\BH} \right) - \frac{ 2\mathcal{M_{\rm d}} m_\BH}{b^2 - m_\BH^2}\,,\\
    &\lambda^{(0,n + 1)}_\text{\tiny int} =  \lambda^{(0,n)}_\text{\tiny int} + \frac{b}{2(n + 1)} \frac{\partial}{\partial b}\lambda^{(0,n)}_\text{\tiny int} \,, \\
    &\frac{(2m + 1)(2n + 3)}{2m + 2n + 3} \lambda_\text{\tiny int}^{(m+1, n)} \nonumber\\
    &=\lambda_\text{\tiny int}^{(m, n)} + \frac{4m(n+1)}{2m + 2n + 3} \lambda_\text{\tiny int}^{(m, n+1)} - b\frac{\partial}{\partial b} \lambda_\text{\tiny int}^{(m, n)}\,,
\end{align}
in terms of the radii $R_\pm = \sqrt{\rho^2+(|z| \mp m_\BH)^2}$.

Within this assumption, we can write down the metric of the SBH-disk model in the approximated form
\begin{align}
\label{bhdiskmetric}
g_{tt}(r,\theta)&\approx-f(r)\left(1+2\nu_{\text{\tiny disk}}\right)\,,\nonumber\\ g_{rr}(r,\theta)&\approx \frac{1}{f(r)}\left(1+2\lambda_{\text{\tiny int}}-2\nu_{\text{\tiny disk}}\right)\,,\nonumber\\
g_{\theta\theta}(r,\theta)&\approx r^2\left(1+2\lambda_{\text{\tiny int}}-2\nu_{\text{\tiny disk}}\right)\,,\nonumber\\ g_{\varphi\varphi}(r,\theta)&\approx r^2\sin^2\theta\left(1-2\nu_{\text{\tiny disk}}\right)\,,
\end{align}
where the metric coefficients are a function of both the radial $r$ and angular $\theta$ coordinates. This metric falls within the category of deformed Schwarzschild space-times~\cite{Chen:2022ynz}.

To perform a further simplification, one can Taylor expand the metric functions in terms of $x = \cos \theta$ as
\begin{equation}
\nu_{\text{\tiny disk}}=\epsilon \mathcal{V}_j(r)|x^j|\,, \quad \lambda_{\text{\tiny int}}=\epsilon \mathcal{L}_j(r)|x^j|\,.
\end{equation}
Each term is weighted by functions of $r$, $\mathcal{V}_j(r)$ and $\mathcal{L}_j(r)$, which depend on $m$, $n$, and $b$ but are independent of $\epsilon$, and has an absolute value  to preserve the equatorial reflection symmetry, and a summation over $j =0, \dots ,j_t$ is implicitly imposed.
This simplification  will allow us to decouple the radial and latitudinal sectors of perturbation equations on the SBH background.

\section{Tidal deformability}
\label{sec: TLN}
\noindent
Nonspinning BHs with axially symmetric thin accretion disks, situated in an external stationary tidal field, will undergo a deformation and develop a multipolar structure due to this external influence. This effect is particularly relevant in coalescing binary systems, where each component exerts tidal forces on its companion. The assumption of a stationary field is valid only during the early inspiral phase, characterized by a significant orbital separation and a slowly varying orbit.

For simplicity, we will consider scalar tidal perturbations.
While the analysis of TLNs in waveform models requires linearized gravitational equations, considering scalar perturbations still allows us to address interesting issues, such as the functional dependence of the TLNs on the disk parameters, without having to deal with the computational complexity of the gravitational case. In this section we provide the explicit form of the perturbation equations and TLNs for scalar perturbations, leaving to Appendix~\ref{appendixA} the discussion of spin-1 tidal fields, which provide similar results. 
Both cases are expected to share the same functional dependence of the gravitational odd-parity sector, ${\rm TLN}_\text{\tiny odd} \propto b^4$, in terms of the physical scale $b$ dictating the geometry of the dressed BH system. As shown in Refs.~\cite{Cardoso:2017cfl,DeLuca:2021ite,Cardoso:2019upw,Arana:2024kaz}, we expect the gravitational even-parity sector to be characterised by a higher power law dependence, ${\rm TLN}_\text{\tiny even} \propto b^5$, possibly because of the direct coupling of the tidal perturbations to those of the matter sector. This motivates the investigation of the measurability of the spin-2 even-parity TLNs for these systems in Appendix~\ref{appendixB}, compared to the one of the scalar tidal deformations shown in Sec.~\ref{sec: fisher}.

The TLNs can be extracted by asymptotically expanding the relevant fields at spatial infinity. In asymptotically Cartesian mass-centered coordinates, the expansion reads~\cite{Cardoso:2017cfl}
\begin{align}
\label{scalarexp}
\phi&=\phi_0 +\sum_{l\geq 1} \left(\frac{1}{r^{l+1}}\left[\sqrt{\frac{4\pi}{2l+1}}\phi_l Y^{l0} + (l'<l \,\text{pole})\right] \right. \nonumber \\
& \left. - \frac{1}{l(l-1)} r^l \left[\mathcal{E}_l+\left(l'< l\,\text{pole}\right)\right]\right)\,,
\end{align}
in terms of the scalar  multipole moment $\phi_l$, tidal field $\mathcal{E}_l$, and spherical harmonics $Y^{l0}$. From this expansion,
it is clear that the computation of the TLNs is based on
the separation between the radially decaying multipolar
response and the external growing
solution. Using the approximation that the multipolar deformation is linear with respect to the strength of the external tide, one can define the TLNs as the ratio of the induced multipole moments to the tidal ones as
\begin{equation}
k_{l}\equiv -\frac{l(l-1)}{r_\BH^{2l+1}}\sqrt{\frac{4\pi}{2l+1}}  \frac{\phi_l}{\mathcal{E}_l}\,.
\end{equation}
The TLNs defined above can be computed by solving the corresponding perturbation equations, requiring regularity at the BH horizon, and finally matching the solution with the asymptotic expansion. This is discussed below for the scalar case.

\subsection{Spin-0 tidal Love number}
\label{spin0TLN}
\noindent
The computation of static TLNs induced by an external scalar perturbation requires solving the massless Klein-Gordon equation
\begin{equation}
\Box\phi=0\,.\label{KGeq}
\end{equation}
Following the projection method of Ref.~\cite{Chen:2022ynz} (see also~\cite{Pani:2013pma}), based on working at leading order in the small deformation parameter $\epsilon \ll 1$, one can separate the radial and the latitudinal sectors of the wave equation. The radial equation can then be recast into a Schr\"odinger-like form for the Fourier modes of the scalar field $\Phi_{l,m_z}(r)$ as 
\begin{equation}
( \partial_{r_*}^2 - V_{\text{\tiny eff}}) \Phi_{l,m_z}(r)=0\,,
\label{kgfoier}
\end{equation}
where $l$ and $m_z$ denote the multipole and azimuthal numbers. 
The effective potential $V_{\text{\tiny eff}}(r)$ can be written as
\begin{widetext}
\begin{align}
V_{\text{\tiny eff}}(r) & = V_{\text{\tiny Schw}}(r) + \epsilon V_{\text{\tiny disk}}(r)  = l(l+1)\frac{f(r)}{r^2} + \frac{f(r)}{r}\frac{df}{dr}\nonumber\\&+\epsilon\left\{ \frac{f(r)}{r}\frac{df}{dr} b_{lm_z}^j\left(4\mathcal{V}_j(r)-2\mathcal{L}_j(r)\right) + \frac{f(r)}{r^2}\left[4a_{lm_z}^j\mathcal{V}_j(r)-c_{lm_z}^j\left(4\mathcal{V}_j(r)-2\mathcal{L}_j(r)\right)\right]-\frac{b_{lm_z}^j}{2}\frac{d^2}{dr_*^2}\left[2\mathcal{V}_j(r)-\mathcal{L}_j(r)\right]\right\}\,,\label{veff2}
\end{align}
\end{widetext}
in terms of the set of coefficients 
\begin{align}
a_{l{m_z}}^j&=\frac{2{m_z}^2}{\mathcal{N}_{l{m_z}}}\int_0^1\frac{x^j\left(P_l^{m_z}\right)^2}{1-x^2}{\rm d}x\,,\label{coeffa}\\
b_{l{m_z}}^j&=\frac{2}{\mathcal{N}_{l{m_z}}}\int_0^1x^j\left(P_l^{m_z}\right)^2{\rm d}x\,,\label{coeffb}\\
c_{l{m_z}}^j&=\frac{2}{\mathcal{N}_{l{m_z}}}\int_0^1x^jP_l^{m_z}\left[\left(1-x^2\right)\partial_x^2-2x\partial_x\right]P_l^{m_z}{\rm d}x\,,\label{coeffc}\\
d_{l{m_z}}^j&=\frac{2}{\mathcal{N}_{l{m_z}}}\int_0^1P_l^{m_z}\left(1-x^2\right)\left(\partial_xx^j\right)\left(\partial_x P_l^{m_z}\right){\rm d}x\,,\label{coeffd}
\end{align}
where the normalization constant $\mathcal{N}_{l{m_z}}\equiv 2(l+m_z)!/[(2l+1)(l-m_z)!]$ is determined by the orthogonality condition
\begin{equation}
\int_{-1}^1P_l^{m_z}(x)P_k^{m_z}(x){\rm d}x =\mathcal{N}_{l{m_z}}\delta_{lk}\,,
\end{equation}
while the tortoise coordinate $r_*$ is defined as
\begin{equation}
\frac{{\rm d}r}{{\rm d}r_*}=f(r)\left\{1+\epsilon b_{lm_z}^j\left[2\mathcal{V}_j(r)-\mathcal{L}_j(r)\right]\right\}\,.\label{tortoiser2}
\end{equation}
In the above equations, the summations over $j$ are implicitly assumed, and we have used the associated Legendre functions $P_l^{m_z}(x)$  to define the corrections to the zeroth-order equation induced by the  deformations of $\mathcal{O}(\epsilon)$ in the metric. While $\epsilon=0$ recovers the pure Schwarzschild result, with the azimuthal numbers $m_z$ being degenerate, in the presence of deformations the spacetime is no longer spherically symmetric and the degeneracy among $m_z$ splits.

To obtain the full expression of the effective potential, one has to compute each coefficient of the sum over the index $j$, which runs until a convergence value $j_t$. 
In the context of quasinormal modes, it was shown that $j_t = 4$ represents already a good degree of convergence~\cite{Chen:2023akf}.
To provide an explicit example, let us focus first on the case $j_t = 0$, for which the angular dependence on $\theta$ vanishes in the metric functions (but their non-spherical form still implies a nondegenerate $m_z$), and the effective potential takes the form (assuming in the following $l = m_z= 2$ and omitting its explicit dependence)
\begin{widetext}
\begin{align}
   & V_{\text{\tiny eff}}^{j_t = 0}  = 
\frac{(r-r_\BH)}{8 r^4} \left\{8 (6 r+r_\BH)-\epsilon \left[ \frac{  \tilde{b}^2 r^2 r_\BH^5 \left(96 \sqrt{\tilde{b}^2 r_\BH^2+r^2-r r_\BH}+\left(73-8 \tilde{b}^2 \left(328 \tilde{b}^4-240 \tilde{b}^2+41\right)\right) r_\BH\right)}{\left(1-4 \tilde{b}^2\right)^2 \left(\tilde{b}^2 r_\BH^2+r^2-r r_\BH\right)^{7/2}} \right. \right. \nonumber \\
& \left. \left. + \frac{8 r^7 \left(4 \sqrt{\tilde{b}^2 r_\BH^2+r^2-r r_\BH}+\left(176 \tilde{b}^4-88 \tilde{b}^2+21\right) r_\BH\right)}{\left(1-4 \tilde{b}^2\right)^2 \left(\tilde{b}^2 r_\BH^2+r^2-r r_\BH\right)^{7/2}}
\right. \right. \nonumber \\
& \left. \left. -\frac{2 r^6 r_\BH \left(32 \sqrt{\tilde{b}^2 r_\BH^2+r^2-r r_\BH}+\left(1936 \tilde{b}^4-912 \tilde{b}^2+135\right) r_\BH\right)}{\left(1-4 \tilde{b}^2\right)^2 \left(\tilde{b}^2 r_\BH^2+r^2-r r_\BH\right)^{7/2}}
\right. \right. \nonumber \\
& \left. \left. + \frac{16 \tilde{b}^6 r_\BH^7 \left(8 \tilde{b}^4 r_\BH+2 \sqrt{\tilde{b}^2 r_\BH^2+r^2-r r_\BH}-2 \tilde{b}^2 r_\BH+r_\BH\right)}{\left(1-4 \tilde{b}^2\right)^2 \left(\tilde{b}^2 r_\BH^2+r^2-r r_\BH\right)^{7/2}}
\right. \right. \nonumber \\
& \left. \left. 
+ \frac{2 \tilde{b}^4 r r_\BH^6 \left(16 \left(\tilde{b}^2-3\right) \sqrt{\tilde{b}^2 r_\BH^2+r^2-r r_\BH}+\left(384 \tilde{b}^6-448 \tilde{b}^4+108 \tilde{b}^2-31\right) r_\BH\right)}{\left(1-4 \tilde{b}^2\right)^2 \left(\tilde{b}^2 r_\BH^2+r^2-r r_\BH\right)^{7/2}}
\right. \right. \nonumber \\
& \left. \left. 
+ \frac{2 r^5 r_\BH^2 \left(48 \tilde{b}^2 \sqrt{\tilde{b}^2 r_\BH^2+r^2-r r_\BH}+\left(2080 \tilde{b}^6+496 \tilde{b}^4-554 \tilde{b}^2+61\right) r_\BH\right)}{\left(1-4 \tilde{b}^2\right)^2 \left(\tilde{b}^2 r_\BH^2+r^2-r r_\BH\right)^{7/2}}
\right. \right. \nonumber \\
& \left. \left. 
+\frac{r^4 r_\BH^3 \left(32 \left(2-3 \tilde{b}^2\right) \sqrt{\tilde{b}^2 r_\BH^2+r^2-r r_\BH}+\left(-7920 \tilde{b}^6+3704 \tilde{b}^4-367 \tilde{b}^2+54\right) r_\BH\right)}{\left(1-4 \tilde{b}^2\right)^2 \left(\tilde{b}^2 r_\BH^2+r^2-r r_\BH\right)^{7/2}}
\right. \right. \nonumber \\
& \left. \left. 
+ \frac{2 r^3 r_\BH^4 \left(16 \left(3 \tilde{b}^4-3 \tilde{b}^2-1\right) \sqrt{\tilde{b}^2 r_\BH^2+r^2-r r_\BH}+\left(1520 \tilde{b}^8+824 \tilde{b}^6-921 \tilde{b}^4+141 \tilde{b}^2-21\right) r_\BH\right)-32 r^8}{\left(1-4 \tilde{b}^2\right)^2 \left(\tilde{b}^2 r_\BH^2+r^2-r r_\BH\right)^{7/2}} \right]\right\}\,.
\end{align}
\end{widetext}
By inspecting this expression, one may immediately realise that it is flattened in the presence of the disk, and that increasing the value of $\tilde{b} = b/r_\BH$ lowers its density and moves it further away from the BH. Furthermore, as expected, the effective potential reduces to the standard Schwarzschild expression both near the horizon and at spatial infinity.

The analytical and manageable expression of the effective potential for $j_t = 0$ allows for a direct analytical computation of the scalar TLNs. In order to do so, we will perform a perturbative approach in the parameter $\epsilon$, such that the equations of motion up to first order take the form  
\begin{align}
& (\partial_{r_*}^2 - V_{\text{\tiny Schw}}) \Phi^{(0)}= 0\,, \nonumber \\
& (\partial_{r_*}^2 - V_{\text{\tiny Schw}})\Phi^{(1)}= \epsilon \, \mathcal{S}[\Phi^{(0)}]\,,
\end{align}
where the tortoise derivative is associated only to the Schwarzschild background, and we have identified a source term $ \mathcal{S}$ which depends on the solution at lower order.

\begin{figure*}[t]
\centering
\includegraphics[width=0.49\textwidth]{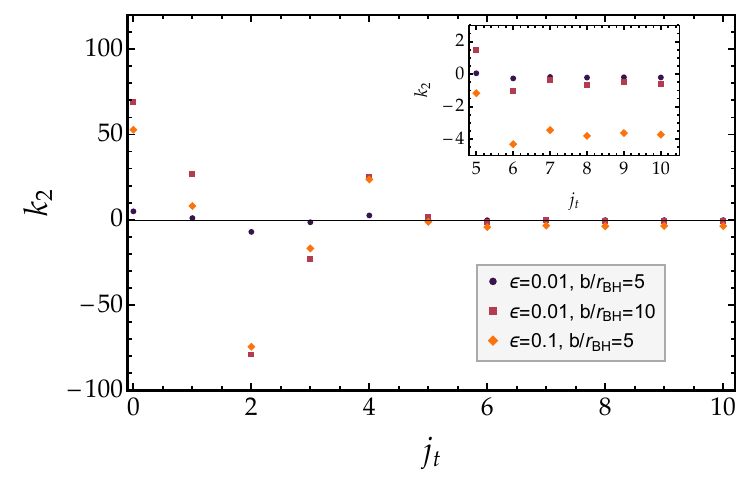}
\includegraphics[width=0.49\textwidth]{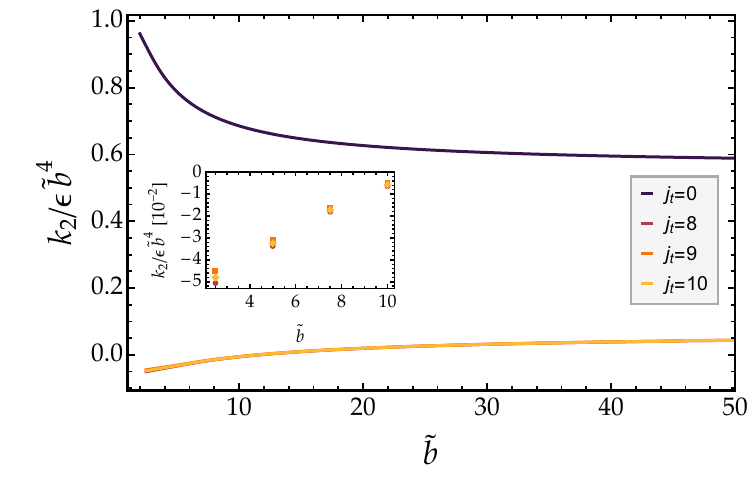}
	\caption{\it Left panel: convergence of the TLN as a function of $j_t$ for different values of $\epsilon$ and $\tilde{b} = b/r_\BH$. From $j_t>5$ on, zoomed in the inset, the TLN converges rapidly to a nonvanishing solution. 
    Right panel: scaling of the TLN with the peak disk position, for various values of $j_t$. The inset zooms over the behaviour at small $\tilde{b}$. For high $\tilde{b}$, the scaling is $k_2 \propto \tilde{b}^4$.}
	\label{fig: TLN-convergence}
\end{figure*}

At zeroth order in $\epsilon$, the solution which is regular at the BH horizon reads
\begin{equation}
\Phi^{(0)} (r) = c_1 \lp \frac{r}{r_\BH} \rp \left[6 \left(\frac{r}{r_\BH}\right)^2-6 \lp \frac{r}{r_\BH} \rp +1\right]\,.
\end{equation}
The source term for $j_t = 0$, once the zeroth-order solution is plugged in, is given by the involved expression
\begin{widetext}
\begin{align}
\mathcal{S}[\Phi^{(0)}] & = \frac{c_1 (r-r_\BH) }{8 r_\BH^3 r^2 \left(1-4 \tilde{b}^2 \right)^2 \sqrt{\tilde{b}^2 r_\BH^2+r (r-r_\BH)}\left(\tilde{b}^2 r_\BH^2+r^2-r r_\BH\right)^3}
\left\{-20 r^7 r_\BH \left[192 \sqrt{\tilde{b}^2 r_\BH^2+r^2-r r_\BH} \right. \right. \nonumber \\
& \left. \left. +\left(48 \left(6-5 \tilde{b}^2\right) \tilde{b}^2+371\right) r_\BH\right]+2 \tilde{b}^4 r_\BH^8 \left[96 \tilde{b}^4 r_\BH+80 \tilde{b}^2 \sqrt{\tilde{b}^2 r_\BH^2+r^2-r r_\BH}+12 \tilde{b}^2 r_\BH+r_\BH\right] \right. \nonumber \\
& \left. -\tilde{b}^2 r r_\BH^7 \left[480 \left(2 \tilde{b}^2+1\right) \tilde{b}^2 \sqrt{\tilde{b}^2 r_\BH^2+r^2-r r_\BH}+\left(4 \tilde{b}^2 \left(496 \tilde{b}^4+28 \tilde{b}^2+73\right)-11\right) r_\BH\right] \right. \nonumber \\
& \left. +192 r^8 \left[5 \sqrt{\tilde{b}^2 r_\BH^2+r^2-r r_\BH}+\left(-8 \tilde{b}^4+4 \tilde{b}^2+22\right) r_\BH\right]+16 r^6 r_\BH^2 \left[10 \left(18 \tilde{b}^2+37\right) \sqrt{\tilde{b}^2 r_\BH^2+r^2-r r_\BH}
\right. \right. \nonumber  \\
& \left. \left. +\left(-192 \tilde{b}^6-204 \tilde{b}^4+873 \tilde{b}^2+410\right) r_\BH\right]-2 r^5 r_\BH^3 \left[2160 \left(2 \tilde{b}^2+1\right) \sqrt{\tilde{b}^2 r_\BH^2+r^2-r r_\BH} \right. \right. \nonumber  \\
& \left. \left. +\left(-3408 \tilde{b}^6+3320 \tilde{b}^4+7659 \tilde{b}^2+1501\right) r_\BH\right]+2 r^2 r_\BH^6 \left[240 \left(2 \tilde{b}^4+7 \tilde{b}^2+1\right) \tilde{b}^2 \sqrt{\tilde{b}^2 r_\BH^2+r^2-r r_\BH} \right. \right. \nonumber  \\
& \left. \left. +\left(2736 \tilde{b}^8+64 \tilde{b}^6+1441 \tilde{b}^4+24 \tilde{b}^2+1\right) r_\BH\right]+2 r^4 r_\BH^4 \left[240 \left(\tilde{b}^2+3\right) \left(6 \tilde{b}^2+1\right) \sqrt{\tilde{b}^2 r_\BH^2+r^2-r r_\BH} \right. \right. \nonumber  \\
& \left. \left. +\left(672 \tilde{b}^8-2496 \tilde{b}^6+6850 \tilde{b}^4+4001 \tilde{b}^2+323\right) r_\BH\right]-r^3 r_\BH^5 \left[160 \left(36 \tilde{b}^4+24 \tilde{b}^2+1\right) \sqrt{\tilde{b}^2 r_\BH^2+r^2-r r_\BH} \right. \right. \nonumber  \\
& \left. \left. +\left(5184 \tilde{b}^8-1168 \tilde{b}^6+9652 \tilde{b}^4+1719 \tilde{b}^2+50\right) r_\BH\right]-960 r^9\right\}
\,,
\end{align}
\end{widetext}
from which the first order solution can be obtained analytically, though we omit its lengthy expression as it provides little additional insight. 
By expanding the total regular solution at spatial infinity, one gets
\begin{widetext}
\begin{align}
& \Phi^{(1)} (r) \approx 6 c_1 \lp \frac{r}{r_\BH} \rp^3 \left\{ 1 + \epsilon \left[ \frac{3840 \tilde{b}^6-1984 \tilde{b}^4+288 \tilde{b}^2+\left(3840 \tilde{b}^7-2304 \tilde{b}^5+440 \tilde{b}^3-30 \tilde{b}\right) \log \left(\frac{(2\tilde{b}-1) r}{r_\BH}\right) }{\left(1-4 \tilde{b}^2\right)^2} \right. \right. \nonumber \\
& \left. \left. + \frac{\left(-3840 \tilde{b}^7+2304 \tilde{b}^5-440 \tilde{b}^3+30 \tilde{b}\right) \log \left(\frac{(2 b+1) r}{r_\BH}\right)-7-\log (16)}{\left(1-4 \tilde{b}^2\right)^2} \right] + \dots  
\right\}  \nonumber \\
&  + c_1 \epsilon 
\frac{\left(25920 \tilde{b}^6+87360 \tilde{b}^5+20880 \tilde{b}^4-37824 \tilde{b}^3-14964 \tilde{b}^2+1532 \tilde{b}+863\right)}{1920 (2 \tilde{b}+1)^2}
\lp \frac{r}{r_\BH} \rp^{-2} + \mathcal{O}(r^{-3}) 
\,,
\end{align}
\end{widetext}
where the dots indicate terms scaling with lower powers of $(r/r_\BH)$, which are irrelevant for the extraction of the tidal response.
By recognizing the growing and decaying behavior as shown in Eq.~\eqref{scalarexp}, one can determine the spin-0 TLN as
\begin{widetext}
\begin{equation}
\label{spin0TLNjt=0}
k_2 = \frac{\epsilon}{11520} \frac{\left(25920 \tilde{b}^6+87360 \tilde{b}^5+20880 \tilde{b}^4-37824 \tilde{b}^3-14964 \tilde{b}^2+1532 \tilde{b}+863\right)}{(2 \tilde{b}+1)^2}
\,.
\end{equation}
\end{widetext}
As expected, the TLN is linear in the disk perturbation parameter $\epsilon$, and it has a leading behaviour $\sim \tilde{b}^4$ for disks peaked at large distances from the BH radius\footnote{Notice that choosing different values of the parameters $m$ and $n$ would impact the coefficients, but not the functional form, of the TLNs.}.

Since an analytical solution is viable only for $j_t = 0$, higher-order corrections at larger $j_t$ demand a numerical approach. In order to compute the corresponding TLNs, we rely on a matching procedure. First, we numerically solve Eq.~\eqref{kgfoier} by imposing regularity of the solution and its derivative at the BH horizon. We further assume the solution at infinity to be described by the following polynomial series:
\begin{align}
\label{infinityansatz}
    &\Phi_\infty (r) \approx  A \lp \frac{r}{r_\BH} \rp^3 \left[ 1+A_1 \lp \frac{r_\BH}{r} \rp + A_2 \lp \frac{r_\BH}{r} \rp^{2}  \right. \nonumber \\
    & \left. +A_3 \lp \frac{r_\BH}{r} \rp^{3} +A_4 \lp \frac{r_\BH}{r} \rp^{4} +A_5 \log\lp \frac{r}{r_\BH}\rp \lp \frac{r_\BH}{r} \rp^{5}\right]
    \nonumber \\
    & +B \lp \frac{r}{r_\BH} \rp^{-2}\,,
\end{align}
which includes both the growing and subleading fall-off behaviours at radial infinity, the latter capturing the linear response to the tidal field. 
We then expand Eq.~\eqref{kgfoier} at spatial infinity and plug the ansatz~\eqref{infinityansatz}. This allows to solve order-by-order for the coefficients $A_1, ..,A_5$ in terms of $A$. Finally, we match at large radii the asymptotic solution and its derivative to the ones obtained by numerical integration, in order to solve for the coefficients $\{A,B\}$. To check the robustness of the procedure, we verified that the obtained values are independent of the choice of the matching radius.
An immediate comparison with Eq.~\eqref{scalarexp} ultimately allows computing the TLN.  

Throughout this procedure, we obtain an excellent agreement with the analytic result reported in Eq.~\eqref{spin0TLNjt=0} in the $j_t=0$ case. The result for higher orders is reported in the left panel of Fig.~\ref{fig: TLN-convergence}, where we show the TLN, $k_2$, as a function of $j_t$, for various choices of the parameters $\epsilon,\tilde{b}$. 
For $j_t<5$, the subleading terms of the growing solution $A_1, ..,A_5$ do not fully converge yet, so the TLN itself, described by the subdominant parameter $B$, is subject to strong fluctuations. Nevertheless, from $j_t>5$ on, where $A_1, ..,A_5$ stabilize as $j_t$ increases, the TLN also converges rapidly to a nonvanishing solution, as evident from the figure. It is also worth commenting that, contrarily to the near-horizon physics of the quasinormal modes~\cite{Chen:2023akf}, the TLNs are a property of the tail of the field solution, and as such it may experience stronger fluctuations until the potential has achieved full convergence.

We also verified that, in the convergence domain at high $j_t$, the functional scaling of the TLN with respect to $\tilde{b}$ remains $k_2 \propto \tilde{b}^4$. This can be appreciated in the right panel of Fig.~\ref{fig: TLN-convergence}, where we show the ratio $k_2/(\epsilon \tilde{b}^4)$ as a function of $\tilde{b}$ for $j_t=0,8,9,10$. As clearly visible in the figure, at high $\tilde{b}$, the scaling of the TLN with this parameter is quartic, in agreement with the $j_t=0$ case. One can also appreciate the convergence of the TLN at higher $j_t$.

\subsection{Environmental vs modified gravity effects}
\label{envvsmg}
\noindent
The presence of an environment around binary BHs, induced for example by secular effects such as accretion, unavoidably induces a nonvanishing TLN for these objects. 
This natural effect may contaminate, and even jeopardize, the ability to use tidal effects to test modified gravity theories or the presence of a BH mimicker, which might affect the BH TLNs in a way similar to that induced by an accretion disk.

Various works have investigated the size of modified gravity effects on the BH TLNs~\cite{Cardoso:2017cfl,Cardoso:2018ptl,DeLuca:2022tkm, Barura:2024uog}. To provide an explicit and model independent example, we will assume that the modified gravity theory is characterised by a certain coupling constant $\alpha_\text{\tiny MG}$ and  energy scale $\Lambda_\text{\tiny MG}$.
The construction of a viable effective field theory for gravity will demand to consider combinations of these parameters satisfying $ \mathcal{O}(\alpha_\text{\tiny MG}/G \Lambda_\text{\tiny MG}^2) \ll 1$ (where we have reintroduced the gravitational constant $G$ to highlight the various scales in the problem). 

As shown in models of higher derivative gravity~\cite{DeLuca:2022tkm} or Chern-Simons gravity~\cite{Alexander:2009tp, Cardoso:2017cfl}, the TLNs of BHs would be nonvanishing and given by
\begin{equation}
k_2^\text{\tiny MG} \propto \frac{\alpha_\text{\tiny MG}}{G \Lambda_\text{\tiny MG}^2} \lp \frac{1}{G m_\BH^2} \rp^2\,.
\end{equation}
This result shows that $k_2^\text{\tiny MG} \lesssim 1$ provides an upper bound for the BH TLNs in the context of effective gravity theories.

\begin{figure}[t!]
	\centering
 	\includegraphics[width=0.45\textwidth]{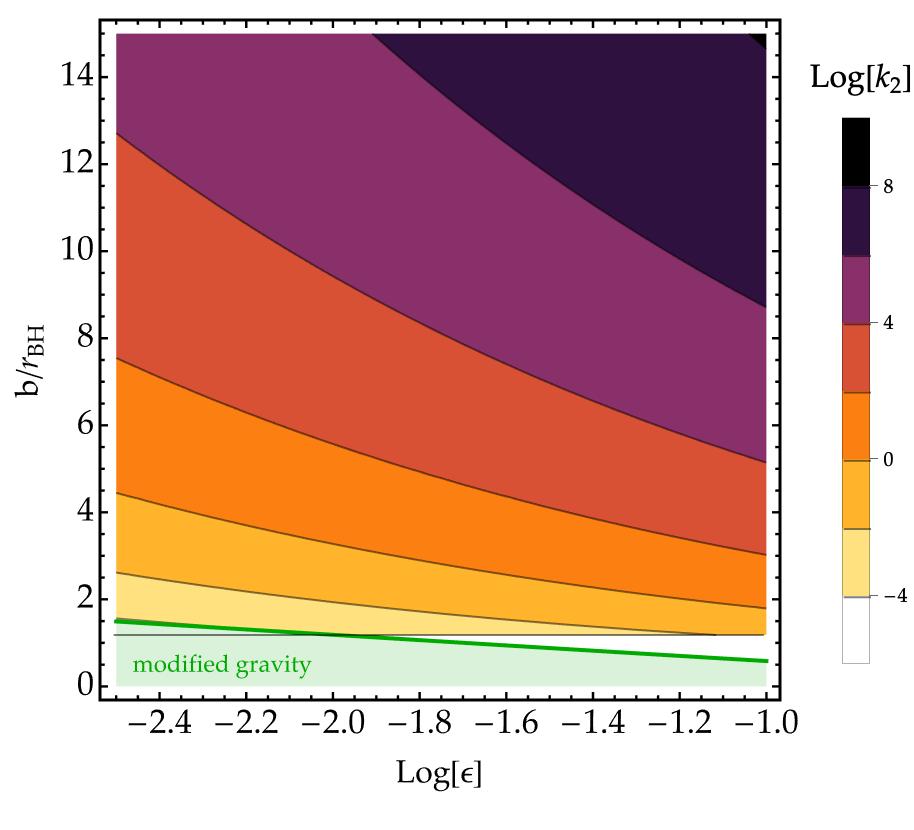}
	\caption{\it Contour plot showing the comparison between the tidal deformabilities of BHs induced by environmental effects, such as thin accretion disks, and modified gravity theories.}
	\label{fig: TLN-MG}
\end{figure}

It is therefore natural to compare their effect on the tidal deformability with the one induced by an external environment, such as a thin accretion disk.
In Fig.~\ref{fig: TLN-MG} we show a contour plot of the TLNs of BHs with accretion disks assuming the simpler $j_t = 0$ case of Eq.~\eqref{spin0TLNjt=0}. The reader may immediately appreciate that values larger than $\mathcal{O}(1)$ are reached whenever $\tilde{b} \gtrsim 5$ and $\epsilon \gtrsim 10^{-2.5}$.  This observation highlights that even milder accretion disks are sufficient to mask the putative presence of modified gravity effects in the theory, represented by the green colored region. In other words, as long as environmental effects are active in the coalescence of binary BHs, their role in tidal deformations could be more important than the one due to modified gravity effects.
An analogous effect would jeopardize tests based on tidal effects for other models, including BH mimickers, as long as their TLN is\footnote{Note that, if $k_2\lesssim{\cal O}(1)$, the 5PN tidal corrections are comparable to the (currently unknown) 5PN point-particle terms, so to test such small tidal effects one would need a  waveform accurate up to at least 5PN order in all terms.} $k_2\lesssim{\cal O}(1)$.

\subsection{Tidal disruption}
\noindent
As shown in the previous section, the TLNs of BHs surrounded by thin accretion disks could be sufficiently large and potentially impact the evolution of coalescing binaries. However, during the final phases of the inspiral, the low-density environment around each object may be unraveled because of the presence of the tidal companion. Similar to mass-shedding events occurring for stars with low compactness, the tidal disruption usually takes place when the binary's semi-major axis is comparable to the Roche radius~\cite{Shapiro:1983du}. For smaller orbital radii, the environment progressively disappears, and the coalescence proceeds with two naked BHs, giving rise to time-dependent TLNs that smoothly approach zero (see Refs.~\cite{DeLuca:2021ite,DeLuca:2022xlz} for a similar tidal evolution in the context of BHs surrounded by bosonic clouds). 

Let us now estimate the Roche frequency associated to the tidal disruption event. For simplicity, we will consider the idealised situation where the accretion disks surrounding each BHs lie in the equatorial plane, over which the merger event is happening. Then, it is easy to find the Roche radius of each BH as
\begin{equation}
r_\text{\tiny Roche} = 2 \gamma m_1 (1+\tilde{b})  \llp \frac{m_2}{m_1 (1+ \epsilon)} \rrp^{1/3}\,,
\end{equation}
where the numerical parameter $\gamma$ takes values
ranging from 1.26 for rigid bodies to 2.44 for fluid ones. We stress that we have assumed the primary object 
to be dominated by the BH itself (neglecting its disk), destroying the disk surrounding its companion. This assumption is consistent since, in our perturbative scheme ($\epsilon\ll1$), the density associated to the disk is much smaller than the one of the BH~\cite{Kotlarik:2022spo,Chen:2023akf}.

The GW frequency associated to tidal disruption for a circular binary then takes the form
\begin{equation}
\label{fcut}
f_\text{\tiny cut} = \frac{1}{2 \sqrt{2} \pi  \gamma ^{3/2} m_1} \sqrt{\frac{(1+\epsilon) (m_1+m_2)}{(1+\tilde{b})^3 m_2}}\,.
\end{equation}
As one can appreciate, accretion disks with wider size (large $\tilde{b}$) or smaller mass (small $\epsilon$) tend to be disrupted earlier during the inspiral. This trend may be compared with the functional dependence of the TLN, which grows with both of them. This implies that, while a small value of $\epsilon$ will make tidal effects less relevant both in size and duration, a small value of $\tilde{b}$ would make TLNs smaller, but their effect in the waveform will last longer.\footnote{Note that the Roche frequency associated with tidal disruption also serves as a useful indicator of when nonlinear TLN effects begin to become significant. Indeed, as shown in Refs.~\cite{Bern:2020uwk, DeLuca:2023mio, Riva:2023rcm}, the quadrupolar response can be expanded as a power law series in terms of the external tidal field induced by the binary companion, and the breaking of linear response should happen around
\begin{equation}
f_\text{\tiny nl}  \sim \sqrt{\frac{k_\text{\tiny l}}{k_\text{\tiny nl}}} \sqrt{\frac{m_1+m_2}{m_2}} \frac{1}{b} \simeq \frac{1}{b} \approx \frac{1}{r_\text{\tiny Roche}}\,,
\end{equation} 
where in the second step we have taken the linear and nonlinear TLNs to be comparable in size, $k_\text{\tiny nl} \simeq k_\text{\tiny l}$, and equal mass binary components, $m_1 \simeq m_2$.}
The interplay between these competing effects will be more transparent in the measurability results of the next section.

\section{Measurability with ET and LISA}
\label{sec: fisher}
\noindent
To assess the prospects of measurability of tidal effects on BHs surrounded by thin accretion disks through GW observations, we will perform a Fisher matrix analysis as described, e.g., in Refs.~\cite{Poisson:1995ef, Vallisneri:2007ev, Cardoso:2017cfl}, which we summarize and show the results of in the next subsections.

\subsection{Fisher matrix analysis}
\noindent
We model the GW signal emitted by the binary using a modified version of the IMRPhenomD waveform model, which encompasses the inspiral, merger, and ringdown phases of coalescence~\cite{Husa:2015iqa,Khan:2015jqa}. The inspiral is divided into an early and a late phase, with the early component described by a post-Newtonian (PN) expansion. In this framework, tidal effects contribute linearly to the point-particle contributions and are fully represented by the TLNs.
In Fourier space, the early part of the signal is expressed as~\cite{Sathyaprakash:1991mt,Damour:2000gg}: 
\be
\tilde h (f) = C_\Omega{\cal A}_\textnormal{PN}  e^{{\rm i} \psi_\text{\tiny PP} (f) + {\rm i} \psi_\text{\tiny Tidal} (f)}\,,
\ee
where $\psi_\text{\tiny PP}$ incorporates terms up to the 3.5PN order~\cite{Damour:2000gg, Arun:2004hn, Buonanno:2009zt, Abdelsalhin:2018reg}, and depends on the binary chirp mass
$\mathcal{M} = (m_1 m_2)^{3/5}/(m_1+m_2)^{1/5}$, the 
symmetric mass ratio $\eta = \eta_1 \eta_2 = m_1 m_2/(m_1+m_2)^2$, 
where $m_{1,2}$ are the component masses, and the (anti)symmetric combinations of the individual spin components: $\chi_{s}=(\chi_{1}+\chi_{2})/2$ 
and $\chi_{a}=(\chi_{1}-\chi_{2})/2$. For the following analysis, we will assume non-spinning binary objects, $\chi_1=\chi_2=0$, compatible with the accretion disk model discussed above, although the spin magnitudes are included in the waveform hyperparameters (see below).
The leading term in the waveform amplitude reads
\be
\mathcal{A}_\textnormal{PN} = \sqrt{\frac{5}{24}} 
\frac{\mathcal{M}^{5/6}f^{-7/6}}{\pi^{2/3}d_L} (1+\textnormal{PN corrections})\,,
\ee
where $d_L$ is the luminosity distance. 
Finally, the geometric coefficient $C_{\Omega}=[F_+^2 (1+\cos^2 \iota)^2 + 4 F_\times^2 \cos \iota]^{1/2}$ is a function of the  
inclination angle $\iota$ between the 
binary line of sight and its orbital angular momentum, 
and on the detector antenna pattern functions $F_{+,\times} 
(\theta, \varphi, \psi)$, which depend on the polarization angle $\psi$ and on the source position in the sky $(\theta, \varphi)$.

The leading tidal correction enters the GW waveform at the
5PN order through $\psi_\text{\tiny Tidal} (f) = -\frac{39}{2}\tilde{\Lambda} (\pi M f)^{5/3}$, in terms of the binary total mass $M=m_1+m_2$ and the effective tidal deformability parameter $\tilde \Lambda$, which depends on the masses and TLNs of each binary component~\cite{Flanagan:2007ix, Vines:2011ud}. We neglect higher-order PN contributions (starting at next order) since they are difficult to measure and increase the dimensionality of the parameter space.

To take into account the possible tidal disruption of  accretion disks during the binary evolution, we will follow the approach proposed in Ref.~\cite{DeLuca:2022xlz} and introduce a frequency-dependent smoothed tidal deformability
\be
\tilde{\Lambda} \to {\cal S} (f) \cdot  \tilde{\Lambda} = \llp \frac{1+e^{-f_\text{\tiny cut}/f_\text{\tiny slope}}}{1+e^{(f-f_\text{\tiny cut})/f_\text{\tiny slope}}} \rrp \cdot \tilde{\Lambda}\,, \label{tapering}
\ee
which allows tidal effects to disappear progressively 
from the characteristic Roche frequency $f_\text{\tiny cut}$, shown in Eq.~\eqref{fcut}, 
with a characteristic slope $f_\text{\tiny slope}$. 
The latter does not impact strongly the results of the analysis, so we will not consider it as a model hyperparameter, and fix it to the value $f_\text{\tiny slope} =f_\text{\tiny cut}/5$.
In our analysis tidal effects are therefore 
described by two quantities, $(\epsilon, \tilde b)$, which determine at the same time the TLNs and Roche frequency, while the overall waveform model depends on 12 
parameters $\vec{\theta}=\{{\cal M},\eta,\chi_s,\chi_a,t_c,\phi_c,
\theta,\phi,\psi,\iota,\epsilon, \tilde b\}$, 
where $(t_c,\phi_c)$ are 
the coalescence time and phase, set to $t_c=\phi_c=0$.
Note that the tapering function~\eqref{tapering} effectively eliminates the tidal-deformability terms from the waveform at frequency $f>f_\text{\tiny cut}$, so the waveform remains the standard IMRPhenomD.\footnote{Let us stress that, even though the TLNs of BH+disk system are expected to be large, they still satisfy the regime of validity of the PN expansion, $\psi_\text{\tiny Tidal} \propto \tilde{\Lambda} (Mf)^{5/3} \lesssim 1$ during the early stages of the inspiral. At larger frequencies, e.g. close to the Roche frequency, the validity of perturbation theory is maintained thanks to the environment's 
disappearance, and thus vanishing TLNs.}

The Fisher-matrix approach~\cite{Poisson:1995ef, Vallisneri:2007ev, Cardoso:2017cfl} is based on the realisation that,  for loud signals with large signal-to-noise ratio, as those expected 
for LISA or third-generation detectors, the posterior distribution of the model hyperparameters $\vec{\theta}$ 
can be described by a multivariate Gaussian distribution centered 
around the {\it true} values $\vec{\hat{\theta}}$, 
with covariance ${\bf \Sigma} = {\bf \Gamma}^{-1}$, where
\be
\Gamma_{ij}= \left\langle \frac{\partial h}{\partial \theta_i}\bigg\vert\frac{\partial h}{\partial \theta_j}\right\rangle_{\vec{\theta}=\vec{\hat{\theta}}}
\ee
is the Fisher information matrix. The 
statistical error on the $i$-th parameter is 
given by $\sigma_i=\Sigma^{1/2}_{ii}$.
In the previous expression, we introduced the scalar product over the detector noise spectral density $S_n(f)$ 
between two waveform templates $h_{1,2}$ as
\be
\langle h_1\vert 
h_2\rangle=4\Re\int_{f_\text{\tiny min}}^{f_\text{\tiny max}} \frac{\tilde{h}_1(f)\tilde{h}^\star_2(f)}{S_n(f)}{\rm d}f \,,
\label{scalprod}
\ee
where $\star$ denotes complex conjugation, such that the signal-to-noise ratio ${\rm snr}=\langle h\vert h\rangle^{1/2}$.
In our analysis we fix the minimum and maximum frequency of 
integration to $f_\text{\tiny min}=1 (10^{-4})$Hz and 
$f_\text{\tiny max}=f_\text{\tiny ISCO}$ for ET and LISA, respectively, where $f_\text{\tiny ISCO}$ is the innermost stable circular orbit
frequency, $f_\text{\tiny ISCO} \simeq 4.4 \, {\rm kHz}\, M_\odot/(m_1+m_2)$.

In the following, we consider optimally-oriented binaries, to remove four angles from the Fisher analysis, which reduces to a $8\times 8$ square matrix. Furthermore, 
we assume 
that binary BHs with thin accretion disks are observed by ET adopting the design ET-D sensitivity curve~\cite{Hild:2010id}, and by LISA adopting the most
optimistic configuration with $2.5 \times 10^6$ km arm-length and an observing time of one year~\cite{LISA:2017pwj, Colpi:2024xhw}.

\begin{figure*}[t!]
	\centering
 	\includegraphics[width=0.325\textwidth]{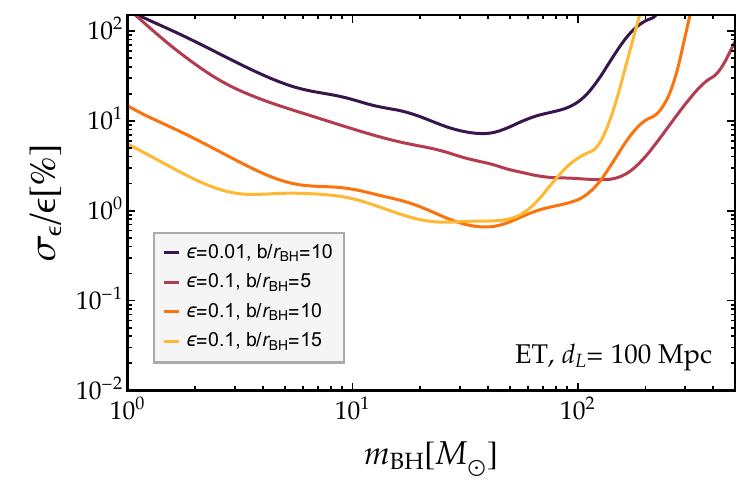}
   	\includegraphics[width=0.325\textwidth]{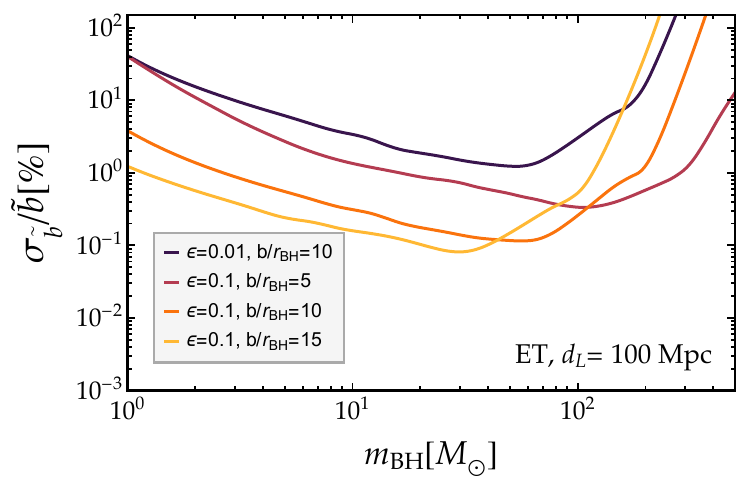}
     \includegraphics[width=0.325\textwidth]{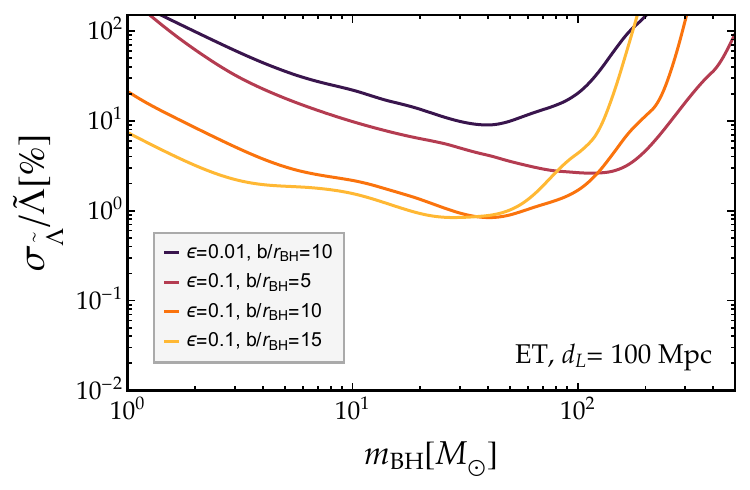}
    \includegraphics[width=0.325\textwidth]{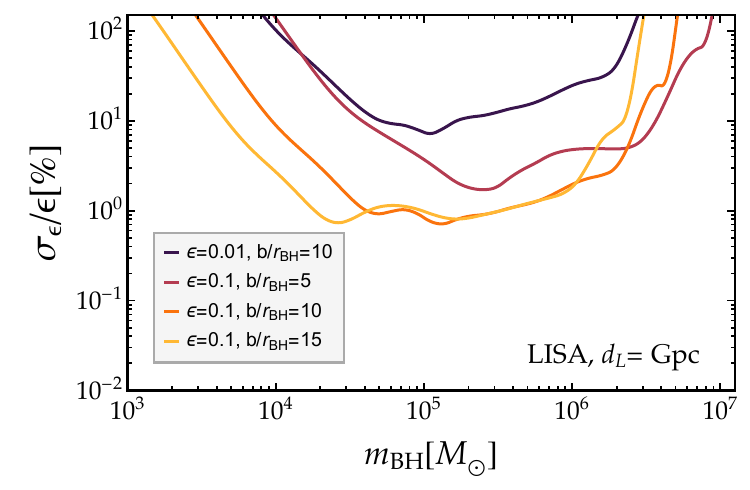}
   	\includegraphics[width=0.325\textwidth]{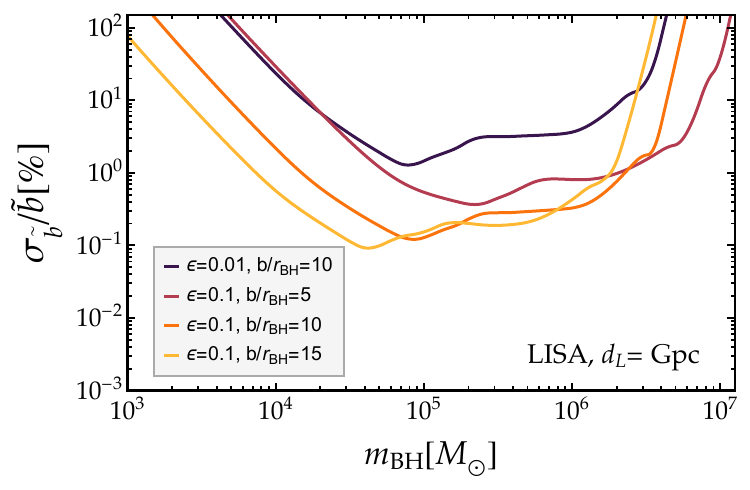}
     \includegraphics[width=0.325\textwidth]{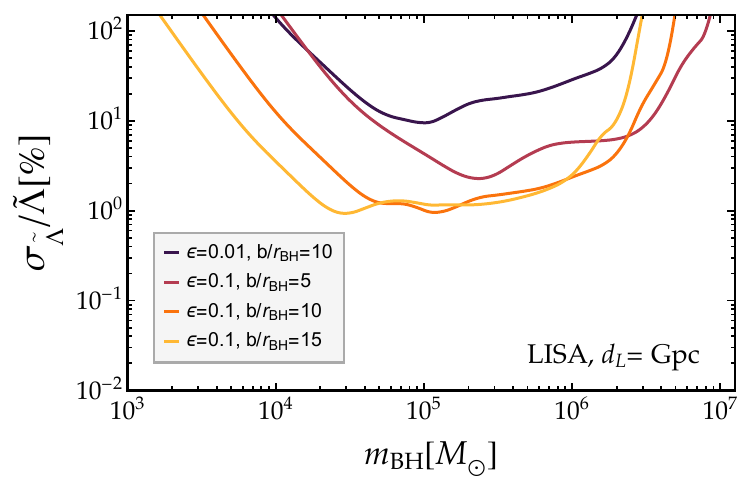}
	\caption{\it (Top row) Relative percentage error on the accretion parameter $\epsilon$ (left), rescaled disk size $\tilde{b}$ (center) and effective TLN $\tilde{\Lambda}$ (right), for a binary BH system with individual masses $m_\BH$ and equal mass ratio, observed by ET at a luminosity distance of $d_L = 100$Mpc. Errors are computed assuming $f_\text{\tiny slope} = f_\text{\tiny cut}/5$. (Bottom row) Same as top panels but considering observable binaries at LISA, with a luminosity distance of $d_L = 1$Gpc.}
	\label{fig: ET-LISA}
\end{figure*}

\subsection{Results}
\noindent
We apply the framework discussed above to investigate the measurability of the tidal parameters for binary BH systems, with each component surrounded by its own thin accretion disk, at Einstein Telescope and LISA, since the sensitivities of current GW detectors still does not allow measuring the tidal parameters with high accuracy. The results of the Fisher analysis are shown in Fig.~\ref{fig: ET-LISA}, for various choices of the disk parameters $\epsilon = (0.01, 0.1)$ and $\tilde{b} = (5,10,15)$. For simplicity we focus on equal mass binaries with $m_1 = m_2 = m_\BH$, although both masses are included as waveform parameters of the Fisher forecast.

The panels of Fig.~\ref{fig: ET-LISA} show that both tidal parameters $(\epsilon, \tilde{b})$ can be measured with high accuracy both at ET and LISA. In particular, the tidal deformability of BH systems with masses in the range $m_\BH \in (1 \divisionsymbol 100)M_\odot$ at distances $d_L = 100$~Mpc will be measured with a relative accuracy of few to ten percent by ET. Similarly, LISA will be able to probe tidal interactions in the whole BH mass range $m_\BH \in (10^3 \divisionsymbol 10^7)M_\odot$ at distances $d_L = 1$~Gpc with a similar precision. The high accuracy in the measurement of these parameters allows reaching percent precision in the measurability of the effective TLN parameter $\tilde{\Lambda}$, which is usually introduced to describe tidal effects in GW waveforms, by properly propagating the relative errors. 

As discussed in the previous section, a smaller value of $\epsilon$ naturally results into worse prospects of measurability for tidal effects, since it would, at the same time, lower the TLN and decrease the Roche frequency~\eqref{fcut}. On the other hand, the trend with the accretion disk parameter $\tilde{b}$ is more complicated, since its larger values increase the TLN but also lower the cut frequency, which manifest as opposite effects in the Fisher analysis. This can be appreciated by looking at the central panels of the plot: for low BH masses, the Roche frequency is high enough to have long-lasting tidal effects in the waveform, such that higher values of $\tilde{b}$ give higher measurability; on the contrary, for larger BH masses, a small $\tilde{b}$ is necessary to have a large enough cut frequency, at the expense of having smaller tidal effects. 

Furthermore, it is worth stressing that our choice for the parameters of the thin accretion disk model, which include the disk mass and size, is also conservative compared to previous results in the literature~\cite{Cardoso:2019upw}. Higher values of $\tilde{b}$ would results into a much better measurability for the tidal parameters in the light portion of the BH mass range.
Finally, following the discussion of modified gravity effects in Sec.~\ref{envvsmg}, it is important to stress that, contrarily to environmental effects, tidal interactions in a modified gravity theory last until the coalescence of the binary system, i.e., they are not characterised by a cut frequency. This could in principle improve their prospects of measurability, even though the corresponding TLNs are much smaller, see Fig.~\ref{fig: TLN-MG}.

\section{Conclusions}
\label{conclusions}
\noindent
Tidal deformability effects are crucial ingredients to properly understand the nature and properties of binary systems we may detect at GW experiments. On this regard, BHs are special, since they are characterised by vanishing TLNs when merging in vacuum. However, the presence of an external environment induces a nonvanishing deformability, which changes the nature of the merger event and may be probed at present and, especially, future GW detectors.

In this work we study the Love numbers of BHs surrounded by nonrotating thin accretion disks of matter. The disk is modeled following the approach of Refs.~\cite{Kotlarik:2018nbd,Kotlarik:2022spo,Chen:2023akf}, based on the existence of an axially symmetric static solution of the Einstein equations, where the disk typical thickness is much smaller than the BH radius. The disk is parameterised by a finite total mass $\mathcal{M}_{\rm d} = \epsilon \, m_\BH$ and, 
while it extents until spatially infinite, its matter content peaks around the scale $b$.

In a perturbative approach in $\epsilon$, we found that the TLNs of these objects, induced by an external test massless scalar perturbation,  are linearly proportional to the disk mass and, assuming the disk's peak to be far enough from the BH horizon, have a leading quartic dependence on $b$. The size of the Love numbers, for a reasonable range of these parameters, highlights that the role of environmental effects in the tidal deformability could be much more important compared to tidal effects induced by modified gravity or BH mimickers.
In particular, we showed that TLNs of BH+disk systems are a possible cause of false general relativity violations in GW observations~\cite{Barausse:2014tra,Gupta:2024gun}.

Finally, we have investigated the measurability of the tidal parameters at future GW experiments like LISA and Einstein Telescope. Both experiments, in the BH mass ranges around the hundreds to millions solar masses, respectively, would be able to measure the model parameters with an accuracy of few to ten percent, within distances around the Gpc. This result depends on the characteristic evolution of the accretion-dressed binary systems which, because of strong tidal interactions around the Roche radius, will lose their environment and continue their evolution in the vacuum until the merger. The positive prospects of detection of the disk's modelling show that the advent of future GW experiments will provide a deep understanding of the environment within which binary BHs evolve.

This work marks the beginning of a comprehensive investigation and offers opportunities for further improvement in several aspects. First of all, one should determine the response of the BH-disk model to spin-2 perturbations  and determine its measurability in comparison to the results obtained assuming the scalar test field as a proxy. Second, it would be interesting to incorporate spin effects in the disk modeling, to estimate their role in the TLN computation and data analysis, or to generalise our results to thick accretion disks. Third, investigating the electromagnetic emission associated to the accretion disks surrounding each binary component could provide a multimessenger signal alongside GWs, offering further insights into the environment of the merger.
Finally, employing a fully Bayesian framework to study  environmental effects at GW experiments would provide a better characterisation of their prospects of measurability.
Exploring these extensions is left to future work.

\section*{Acknowledgments}
\noindent
We thank Richard Brito and Loris Del Grosso for interesting discussions. We also thank Sumanta Chakraborty and Avijit Chowdhury for pointing out a typo in the definition of the radii $R_\pm$ in an earlier version of the manuscript.
E.C. acknowledges financial support provided under the European Union’s H2020 ERC Advanced Grant “Black holes: gravitational engines of discovery” grant agreement no. Gravitas–101052587. Views and opinions expressed are however those of the author only and do not necessarily reflect those of the European Union or the European Research Council. Neither the European Union nor the granting authority can be held responsible for them. E.C. acknowledges support from the Villum Investigator program supported by the VILLUM Foundation (grant no. VIL37766) and the DNRF Chair program (grant no. DNRF162) by the Danish National Research Foundation. This project has received funding from the European Union's Horizon 2020 research and innovation programme under the Marie Sklodowska-Curie grant agreement No 101007855 and No 101131233.
V.DL. is supported by funds provided by the Center for Particle Cosmology at the University of Pennsylvania. 
P.P. is partially supported by the MUR PRIN Grant 2020KR4KN2 ``String Theory as a bridge between Gauge Theories and Quantum Gravity'' and by the MUR FARE programme (GW-NEXT, CUP:~B84I20000100001).

\appendix

\section{Spin-1 tidal Love number}
\label{appendixA}
\noindent
Similarly to what discussed in the main text, one may also compute the tidal deformation of a BH surrounded by a thin accretion disk induced by a spin-1 external tidal field.
Since we aim primarily to investigate the functional dependence of the TLNs on the model parameters, for simplicity we will focus only on the $j_t = 0$ term in the deformed metric of Eq.~\eqref{bhdiskmetric}.

The corresponding TLNs are defined similarly to the scalar case, by performing an asymptotic expansion for the components of vector field, $A_t$ and $A_\varphi$, as~\cite{Cardoso:2017cfl}
\begin{align}
\label{eq:Aphiexpansion}
A_t &= - \frac{Q}{r}+\sum_ {l\geq 1} \left(\frac{2}{r^{l+1}}\left[\sqrt{\frac{4\pi}{2l+1}}Q_l Y^{l0} + \left(l'<l\,\text{pole}\right)\right] \right. \nonumber \\
& \left. -\frac{2}{l\left(l-1\right)}r^l\left[\mathfrak{E}_l Y^{l0}+\left(l'< l\,\text{pole}\right)\right]\right)\,, \nonumber \\
A_\varphi &= \sum_ {l\geq 1} \left(\frac{2}{r^{l}}\left[\sqrt{\frac{4\pi}{2l+1}}\frac{J_l}{l} S_\varphi^{l0} + \left(l'<l\,\text{pole}\right)\right] \right. \nonumber \\
& \left. +\frac{2}{3l\left(l-1\right)}r^{l+1}\left[\mathfrak{B}_l S_\varphi^{l0}+\left(l'< l\,\text{pole}\right)\right]\right)\,,
\end{align}
in terms of the electric and magnetic multipole moments $Q_l, J_l$, the corresponding tidal fields $\mathfrak{E}_l,\mathfrak{B}_l$, and the harmonic function $S_\varphi^{lm_z} = \sin \theta \, \partial_\theta Y^{lm_z}$. The TLNs in the polar and axial sectors are then defined as
\begin{align}
k^{(s=1), {\rm p}}_{l} & \equiv - \frac{l(l-1)}{r_\BH^{2l+1}}\sqrt{\frac{4\pi}{2l+1}}  \frac{Q_l}{\mathfrak{E}_{l}}\,, \nonumber \\
k^{(s=1), {\rm a}}_{l} & \equiv  \frac{3(l-1)}{r_\BH^{2l+1}}\sqrt{\frac{4\pi}{2l+1}}  \frac{J_l}{\mathfrak{B}_{l}}\,.
\label{Lovenumbersdef1}
\end{align}
The equation of motion for a vector field $A_\mu$ in the SBH-disk spacetime is given by
\be
\label{EOMspin1}
\nabla_\nu F^{\mu \nu} = 0\,,
\ee
where we have defined the electromagnetic tensor $F_{\mu \nu} = \partial_\mu A_\nu - \partial_\nu A_\mu$ and the covariant derivative $\nabla_\nu$.
The spin-$1$ field can be decomposed as~\cite{Hui:2020xxx} 
\begin{align}
 A_\mu  & = \sum_{l, m_z} \lp a_0, a_r, a^{(l)}  \partial_\theta, a^{(l)}  \partial_\varphi \rp Y^{lm_z} \nonumber \\
 & + \sum_{l, m_z} \lp 0,0, a^{(T)}/ \sin \theta \, \partial_\varphi, - a^{(T)} \sin \theta \, \partial_\theta \rp Y^{lm_z}
 \,.
\label{fieldsansatz}
\end{align}
The quantities $a_0, a_r, a^{(l)}$ are scalars under SO(3), and have been expanded in spherical harmonics, omitting the $(l,m_z)$ dependence for simplicity. They describe the degrees of freedom in the polar (parity-even) sector, and mix among themselves, while the function $a^{(T)}$ has been obtained by expanding the spin-1 angular components in vector spherical harmonics, and describe the axial (parity-odd) sector. Furthermore, they are all function of the time and radial coordinates. In the following, we will focus on the quadrupolar term $l = 2$, $m_z = 0$.

\begin{figure}[t!]
	\centering
 	\includegraphics[width=0.45\textwidth]{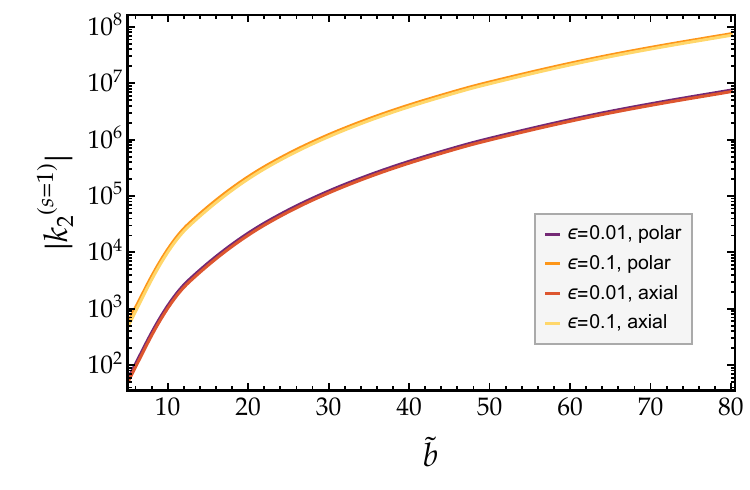}
	\caption{\it Behaviour of the TLN of a Schwarzschild BH surrounded by a thin accretion disk in terms of $\tilde{b}$, induced by both the parity-even (polar) and parity odd (axial) sectors of a spin-1 tidal field, for different values of $\epsilon$. Note that polar and axial TLNs are almost the same.}
	\label{fig: TLN-spin1}
\end{figure}

Since there is only a single degree of freedom in the polar sector, one can first use the gauge freedom to fix $a^{(l)} = 0$. Then, one can introduce a new auxiliary variable $\Psi$, defined as~\cite{Hui:2020xxx}
\begin{equation}
\Psi = \frac{r^2}{\sqrt{6}} (a_0' - \dot a_r)\,,
\end{equation}
where the prime/dot denotes a radial/time derivative, to simplify the system and integrate out both $a_0$ and $a_r$. In particular, in the static limit, $a_r = 0$ and 
\begin{align}
& a_0  = \frac{(r-r_\BH) \Psi'}{\sqrt{6} r} \nonumber \\
& -\epsilon \sqrt{\frac{2}{3}} \frac{ r_\BH   (r_\BH-2r) (r-r_\BH) \left[ \tilde{b}^2 r_\BH^2+2r (r_\BH-r)\right] }{8r \left[\tilde{b}^2 r_\BH^2+r (r-r_\BH)\right]^{5/2}} \Psi\,.
\end{align}
The equation of motion \eqref{EOMspin1} for $\Psi$ then takes the schematic form, up to first order in $\epsilon$,
\begin{align}
& r (r_\BH-r) \Psi^{(0)''}-r_\BH \Psi^{(0)'}+6 \Psi^{(0)} = 0\,, \nonumber \\
&  r (r_\BH-r) \Psi^{(1)''}-r_\BH \Psi^{(1)'}+6 \Psi^{(1)}= \epsilon \, \mathcal{S}[\Psi^{(0)}]\,,
\end{align}
as a function of a source term $\mathcal{S}$, that depends on the solution at lower order.

\begin{figure*}[t!]
	\centering
 	\includegraphics[width=0.49\textwidth]{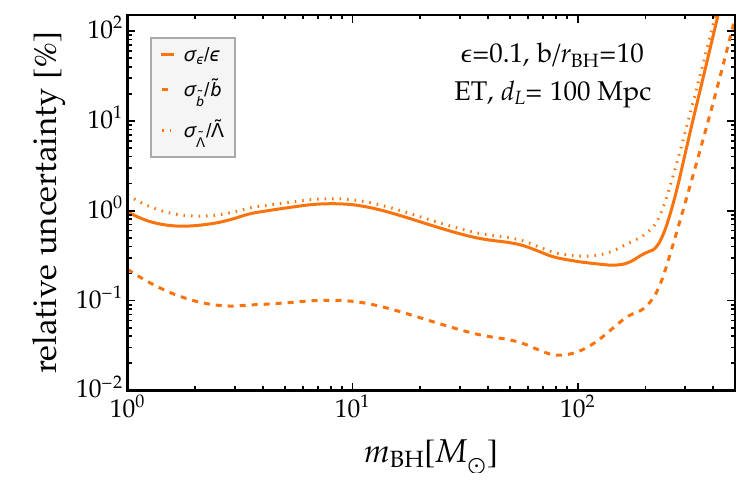}
   	\includegraphics[width=0.49\textwidth]{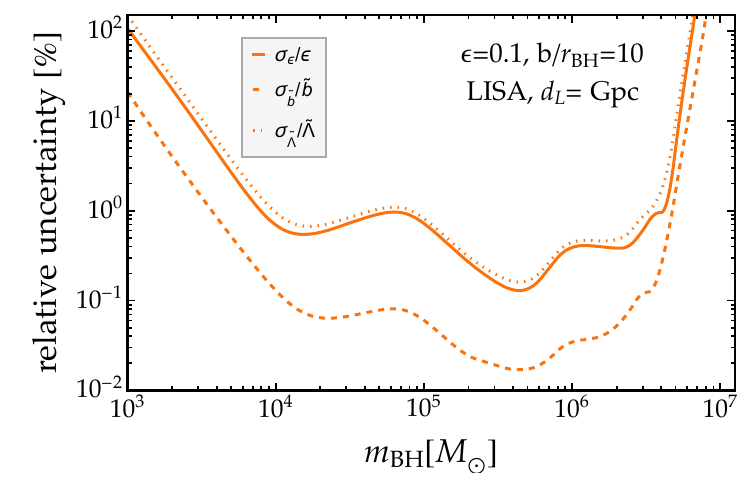}
	\caption{\it Relative percentage error on the accretion parameter $\epsilon$, rescaled disk size $\tilde{b}$ and effective TLN $\tilde{\Lambda}$, for a binary BH system with individual masses $m_\BH$ and equal mass ratio, observed by ET at a luminosity distance of $d_L = 100$Mpc (left panel) and by LISA at luminosity distance of $d_L = 1$Gpc (right panel). Errors are computed assuming $f_\text{\tiny slope} = f_\text{\tiny cut}/5$. In both analyses we have assumed a different scaling for the TLN, $k_2 \propto \tilde{b}^5$, as expected for parity-even gravitational perturbations.}
	\label{fig: ET-LISA-AppB}
\end{figure*}

Similarly to the scalar case, also this equation can be solved perturbatively. After imposing the regularity of the solution at the BH horizon, the solution at zeroth order reads
\be
\Psi^{(0)}= c_1 \lp \frac{r}{r_\BH} \rp^2 \llp \lp \frac{r}{r_\BH} \rp - \frac{3}{4} \rrp\,.
\ee
Once plugged in, the source term at first order in $\epsilon$ takes the involved expression 
\begin{widetext}
\begin{align}
\mathcal{S}[\Psi^{(0)}] & = \frac{c_1}{32} \lp \frac{r}{r_\BH} \rp^2  \left(\tilde{b}^2 r_\BH^2+r^2-r r_\BH\right)^{-7/2} \llp  6 \tilde{b}^4 r_\BH^7+8 \left(26 \tilde{b}^2+17\right) r^5 r_\BH^2-2 \left(284 \tilde{b}^2+29\right) r^4 r_\BH^3 \right.  \nonumber \\
& \left.  +\tilde{b}^2 \left(33-52 \tilde{b}^2\right) r r_\BH^6-4 \left(16 \tilde{b}^4-139 \tilde{b}^2+1\right) r^3 r_\BH^4+\left(108 \tilde{b}^4-229 \tilde{b}^2+6\right) r^2 r_\BH^5+32 r^7-112 r^6 r_\BH \rrp\,.
\end{align}
\end{widetext}
The equation of motion at order $\epsilon$ does not admit an analytical solution, and one has to solve it numerically. To do so, we follow the same approach discussed in Sec.~\ref{spin0TLN} for the scalar case when $j_t > 0$. The result is shown in Fig.~\ref{fig: TLN-spin1}, for different values of $\epsilon$. As one can appreciate, 
the TLN grows linearly in $\epsilon$, and has a characteristic growing behavior with $\tilde{b}$, with slope $\sim \tilde{b}^4$.

In the parity-odd sector, the problem can be similarly addressed by studying the equation of motion for the field $a^{(T)}$ in a perturbative series in $\epsilon$ as 
\begin{align}
 r (r_\BH-r) a^{(T,0)''}-r_\BH a^{(T,0)'} + 6 a^{(T,0)}  & = 0\,, \nonumber \\
  r (r_\BH-r) a^{(T,1)''}-r_\BH a^{(T,1)'} + 6 a^{(T,1)} & = \epsilon \, \mathcal{S}[a^{(T, 0)}]\,,
\end{align}
in terms of a source term $\mathcal{S}[a^{(T,0)}]$, evaluated at the lower order solution. The latter reads, after imposing the condition of regularity at the BH horizon,
\be
a^{(T,0)} = c_1 \lp \frac{r}{r_\BH} \rp^2 \llp \lp \frac{r}{r_\BH} \rp - \frac{3}{4} \rrp\,,
\ee
such that the source becomes 
\begin{widetext}
\begin{align}
\mathcal{S}[a^{(T,0)}] & = c_1 \frac{3  r^2  (r-r_\BH) (r_\BH-2 r)^2 \left(-\tilde{b}^2 r_\BH^2+2 r^2-2 r r_\BH\right)}{8 r_\BH^2 \left(\tilde{b}^2 r_\BH^2+r^2-r r_\BH\right)^{5/2}}\,.
\end{align}   
\end{widetext}
The corresponding TLN is found numerically and shown in Fig.~\ref{fig: TLN-spin1},   with trends and values comparable to the polar case. These results show that the TLNs of BH+disk systems have a functional dependence on the model parameters which is independent from the external applied tidal field.
Lastly, we expect a similar functional behavior also for the odd-parity sector of gravitational perturbations, $k_2^{(s=2, \rm odd)} \propto \tilde{b}^4$, contrarily to the even-parity one discussed in the next Appendix~\cite{Cardoso:2017cfl,DeLuca:2021ite,Cardoso:2019upw,Arana:2024kaz}.

\section{Measurability for spin-2 even tidal Love number}
\label{appendixB}
\noindent
As shown in Refs.~\cite{Cardoso:2017cfl,DeLuca:2021ite,Cardoso:2019upw,Arana:2024kaz}, the even-parity sector of gravitational perturbations is expected to experience a different functional behaviour for the associated TLNs compared to the odd-parity one, possibly because of the coupling of the perturbation modes of the tidal field and matter content. In our framework, we thus expect a different power law of the form $k_2^{(s=2, \rm even)} \propto \epsilon \, \tilde{b}^5$. While a dedicated study on its computation is left to future work, here we would like to investigate the effect of such a different slope in the measurability of TLNs of BH binaries surrounded by disks at ET and LISA.
The results are shown in Fig.~\ref{fig: ET-LISA-AppB} for the representative choice of $\epsilon = 0.1$ and $\tilde{b} = 10$.
It is manifest that, compared to the results of the scalar field, the higher power in $\tilde{b}$ results into an order-of-magnitude improvement in the expected accuracy of measurability of the tidal parameters.

\bibliography{draft}

\end{document}